\documentclass[fleqn,12pt,twoside]{article}
\usepackage{amsmath,amssymb}
\usepackage{amstext}
\usepackage{espcrc1}
\usepackage{graphicx} 
\usepackage{epsfig}
\usepackage{subfigure}

\title
{Spectrometer Calibration by Expectation Maximization Method}
\author{L. Yuan \address[MCSD]{Department of Physics, \\ 
       Hampton University, Hampton, VA 23668}}
\begin{document}

\maketitle

\begin{abstract}
Expectation Maximization (EM) algorithm is a parameter estimation
method from incomplete observations. In this paper, an implementation
of this method to  the calibration of HKS spectrometer at Jefferson Lab
is described. We show that the application of EM method is able to
calibrate the spectrometer properly in the presence of high background
noise, while the traditional nonlinear Least Square method fail. The
preliminary results of HKS spectrometer calibration is presented.
\end{abstract}
 
\section{Introduction}

Expectation Maximization (EM) algorithm is a statistical method for
parameter estimation from incomplete observations. It is an extension
of the Maximum Likelihood (ML) method well
known to physicists. 
This method was first proposed by A. Dempster etc in
\cite{demp}. Since its introduction, this method has been used in a
wide varieties of applications such as signal processing, medical
image processing and genetics, to name a few
(\cite{IEE-em}).

The EM method is well suited to handle problems with observations
diluted by large amount of noise, since it is not known a priori a
observation is signal or noise.  That is why it is introduced into
High Energy physics for track reconstruction in the ATLAS detector at
LHC (\cite{nim-ea}, \cite{cpc-tr}). In the presence of track noise,
the EM based tracking algorithm can obtain a track resolution more
than two order of magnitude better than traditional Least Square
tracking method .

We have used the EM method in the spectrometer calibration for HKS
experiment at Jefferson Lab. Jefferson Lab HKS experiment aims at
obtaining high resolution hypernuclear spectroscopy by (e,e'K)
reaction. To achieve this goal,it is essential to perform a proper
spectrometer calibration to optimize the reconstruction resolution of
the momentum and angles of scattering electrons and Kaons
(\cite{arxiv}). The only high precision calibration method is to make
use of the known masses of $\Lambda$,$\Sigma^0$ hyperons produced from
hydrogen in CH$_2$ target and the narrow width of $^{12}_\Lambda$B
hypernuclear ground state from $^{12}$C target(\cite{PDB}).  These
masses can be produced at the same spectrometer kinematics as the
production of hypernuclei.

\section{Nonlinear Least Square Method}

Let $\{\boldsymbol\pi\}$ denote the set of parameters which defines the
reconstruction function. For example, the $\{\boldsymbol\pi\}$ can be a set of polynomial
coefficients in the polynomial expansion of reconstruction function. The task of calibration now is to find the
best set of parameters $\{\boldsymbol\pi\}$ to optimize the
reconstruction resolution. The missing mass $E_m$ of
(e,e'K) reaction
can be calculated from the focal plane measurement $X_i,\quad
i=1,\ldots N$, $i$ denotes each event number, and the reconstruction
parameters $\{\boldsymbol\pi\}$ : 
\begin{equation}
 E_m^{i} =
\mathfrak{f}(X_i,\{\boldsymbol\pi\}), 
\end{equation} 
where $\mathfrak{f}$ is a nonlinear function. The format of $\mathfrak{f}$ can
be derived from the kinematics equations. $X_i$ represent the
trajectories of the particles at spectrometer focal plane. 
Let $\Delta M_i$ be the difference between the calculated mass and the
known mass value from Particle Data Book $M^{PDB}$,
\begin{equation}
\Delta M_i= E_m^{i} - M^{PDB}. 
\end{equation}
Finally, we define a Chisquare as the weighted sum of squared $\Delta M_i$ over all events:
\begin{equation}
 \chi ^2 = \sum_{i=1}^N w_i\Delta M_i^2 = \sum_{i=1}^N
 w_i(\mathfrak{f}(X_i,\{\boldsymbol\pi\})- M^{PDB})^2,
\label{eq:chis} 
\end{equation}
where $w_i$ is the relative weights of $\Lambda$,$\Sigma$ and
$^{12}_\Lambda$B GS events. 

The set of parameters $\{\boldsymbol\pi\}$ which minimize the Chisquare
function will define our optimized reconstruction function. This is a
typical nonlinear Least Square (NLS) problem. The Chisquare function
 is still a complex nonlinear function and
have to be minimized by numerical method. It is carried out by
using CERNLIB Fortran program package LEAMAX (\cite{leamax}).
 
\begin{figure}
\begin{center}
\includegraphics [width=12cm]{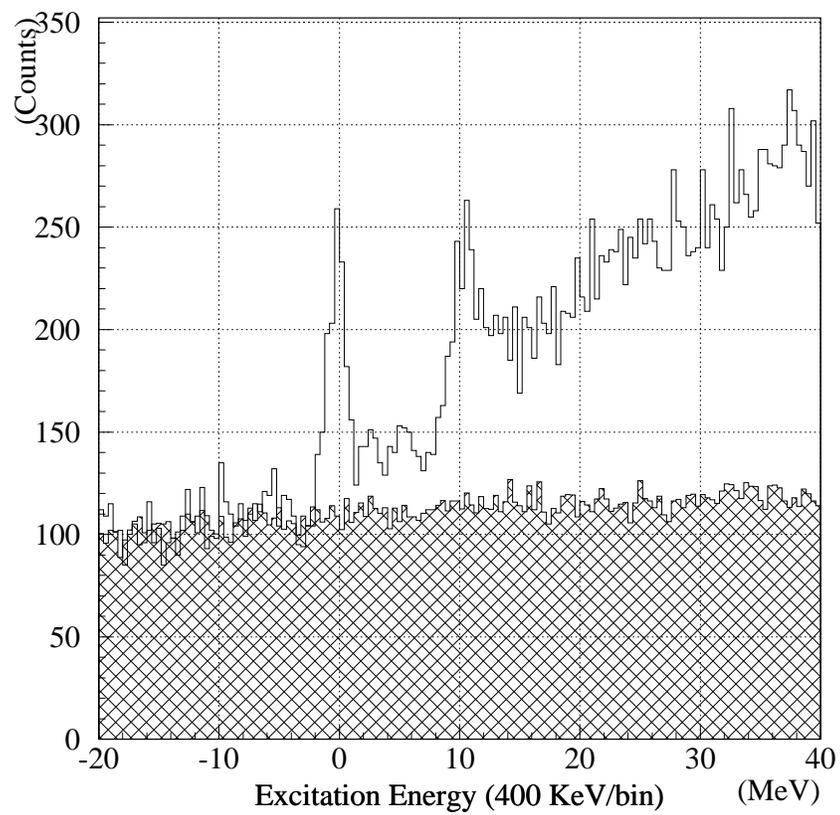}
\end{center}
\caption{$^{12}_\Lambda$B excitation energy spectrum. The shaded region is
  accidental backgroud.}
\label{fig:c12}
\end{figure}

\begin{figure}
\begin{center}
\includegraphics [width=12cm]{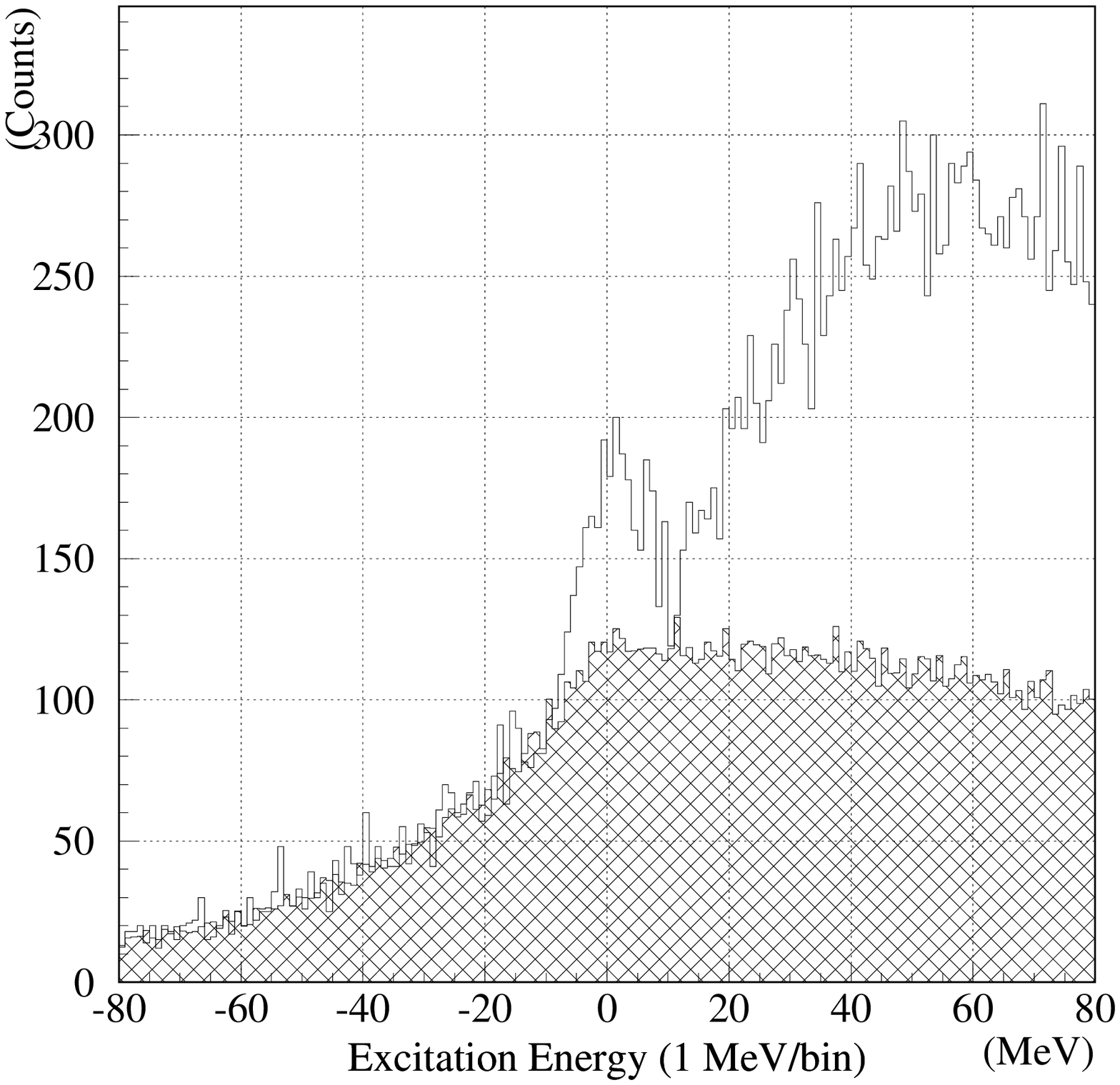}
\end{center}
\caption{$^{12}_\Lambda$B missing mass spectrum used in the improper
  calibration of HKS spectrometer by nonlinear Least Square
  method. The shaded region is  accidental backgroud.}
\label{fig:nls}
\end{figure}

In case we have a clean signal of hyperons and hypernuclear bound
states (The signal to noise (S/N) ratio better than 6:1), the NLS method works
well, as is shown for simulated HKS data in our Arxiv paper (\cite{arxiv}). However, the HKS spectrometer setup detects very
forward angle e' and Kaons in order to increase hypernuclear yield.
Thus in real experiment, we see high accidental background between
Kaon arm and electrons produced by
Bremsstrahlung photons. For the real data, The S/N ratio
in the missing mass spectrum is almost 1:1 (fig.\ref{fig:c12}). Applying
the NLS method to the data results in wrong calibration. As shown in
fig.\ref{fig:nls}, the accidental background under the
$^{12}_\Lambda$B gs peak, which should be flat, now forms an
artificial ``bump'' as a result of the NLS calibration. Clearly, the EM
algorithm which is robust to noise observations should be used for the
spectrometer calibration of HKS.

\section {Expectaton Maximization method}

We have two condiderations in order to implement the EM method:
\begin {enumerate}

\item EM method is used for parameter estimation with incomplete
   observation. Taking advantage of this feature, we can define a
   variable $S_i,\quad
i=1,\ldots N$, which denotes whether event $i$ is a real signal (real
   coincidence) : $S_i=1$
    or noise (accidental coincidence): $S_i=0$, although $S_i$ can not be
   observed by the spectrometer. Thus instead of minimize the
   chisquare in eq.\ref{eq:chis}, we will minimize an energy function
   defined as
\begin  {equation}
E(\{S_i\},\{\boldsymbol\pi\}) = \sum_i [S_i\Delta M_i^2+\lambda(S_i-1)^2],
\end {equation}
where $\lambda$ is a cut off paramter.

\item There is a large number of parameters in the parameter set $\{\boldsymbol\pi\}$, to aviod the
   calibration process ending up in a local minimum, we introduce an
   annealing process (\cite{cpc-tr}). One requires each configuration of the
   system with energy $E$ obey the Bolzmann distribution at temperature
   $T$. One then  minimizes the expectation value of the energy function at
   successively lower temperatures until final result at $T
   \rightarrow 0$.

According to Bolzmann distribution, the probability for the system to
have configuration $\{S_i,\Delta M_i, i=1,\ldots N\}$,  is:
\begin {equation}
P(\{S_i\},\{\boldsymbol\pi\}) = \mathrm e^ {-\beta E(\{S_i\},\{\boldsymbol\pi\})}/Z,
\label{eq:prop}
\end{equation}

where $\beta$ is the inverse temperature $\beta=1/T$ and Z is the partition function:
\begin{equation}
 Z = \sum_{\{S_i\}}\int d\{\boldsymbol\pi\} \mathrm e^ {-\beta
   E(\{S_i\},\{\boldsymbol\pi\})}
\end{equation}
\end {enumerate}

The EM algorithm can be divided into the expectation step and
the maximization step. In the expectation step, the expectaion value of
the energy function over the unobserved variable is calculated:
\begin {equation}
\begin {split}
Q(\{\boldsymbol\pi\} \mid \{\boldsymbol\pi\prime\})&=\sum_{\{S_i\}}
E(\{S_i\},\{\boldsymbol\pi\})P(\{S_i\}\mid \{\boldsymbol\pi\prime\}) \\
&=\sum_{\{S_i\}} E(\{S_i\},\{\boldsymbol\pi\}) \cdot
\frac{P(\{S_i\},\{\boldsymbol\pi\prime\})}{P_M(\{\boldsymbol\pi\prime\})}
\end{split}
\label{eq:exp}
\end {equation}
where $P(\{S_i\}\mid \{\boldsymbol\pi\prime\})$  is the probability function of assignment variables $S_i$
conditioned on the parameter set $\{\boldsymbol\pi\prime\}$,
$P_M$ is the marginal probability function:
\begin {equation}
P_M(\{\boldsymbol\pi\prime\}) = \sum_{\{S_i\}}
P(\{S_i\},\{\boldsymbol\pi\prime\})= \mathrm e^{-\beta E_{eff}}/Z, 
\label{eq:marg}
\end{equation}
Effective energy 
\begin{equation}
E_{eff} = -\frac{1}{\beta} \sum_{i=1}^{N}\log(\mathrm e^{-\beta
  \lambda} + \mathrm e^{-\beta \Delta M_i^2}). 
\end{equation}

\begin{figure}
\begin{center}
\includegraphics [width=12cm]{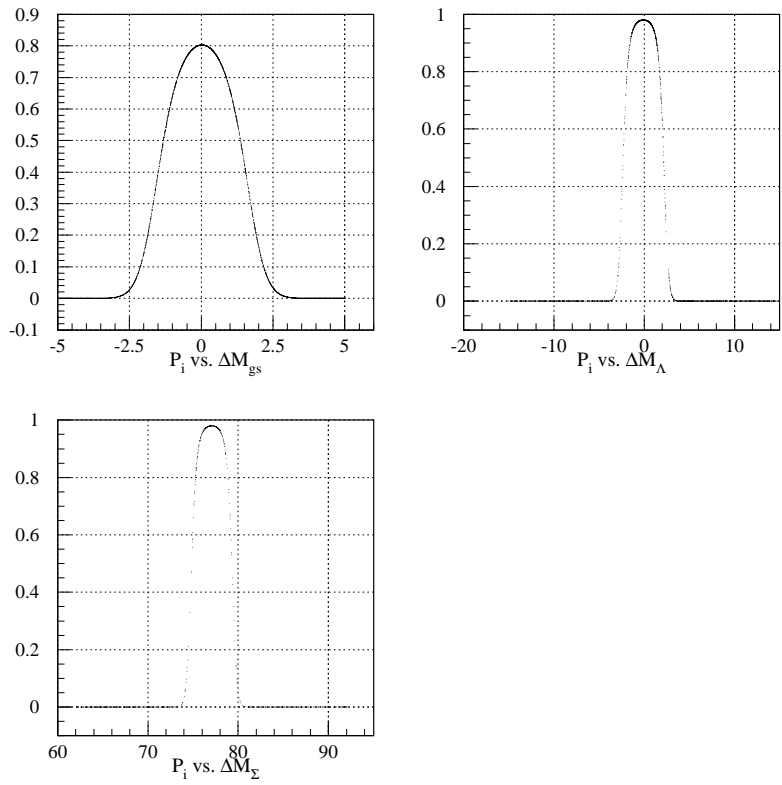}
\end{center}
\caption{The dependences of calculated probability functions $p_i$ for $\Lambda$,$\Sigma$
  and $^{12}_\Lambda$B on mass differences $\Delta M_i$}
\label{fig:rw}
\end{figure}

Substitute equations \ref{eq:prop} and \ref{eq:marg} into equation
\ref{eq:exp}, we can write up the expectation value as:
\begin {equation}
\begin {split}
Q(\{\boldsymbol\pi\} \mid \{\boldsymbol\pi\prime\})& = \sum_{i=1}^{N}
[\Delta M_i^2 \frac{\mathrm e^{-\beta \Delta M_i\prime^2}}{\mathrm e^{-\beta
  \lambda} + \mathrm e^{-\beta \Delta M_i\prime^2}} +\lambda
\frac{\mathrm e^{-\beta \lambda}} {\mathrm e^{-\beta
  \lambda}+ \mathrm e^ {-\beta \Delta M_i\prime^2}}] \\
&= \sum_{i=1}^{N}[\Delta M_i^2 p_i\prime+\lambda p_0\prime].
\end {split}
\label {eq:pi}
\end{equation}
$p_i\prime$ can be interpreted as the probability that event $i$ is a real
signal. The minimization step is then to minimize the $Q$ function
with respect to the parameter set $\{\boldsymbol\pi\}$. Because the second term is independent of $\{\boldsymbol\pi\}$, in the maximization
step, we will minimize function:
\begin {equation}
\mathfrak{g}(\{\boldsymbol\pi\} \mid \{\boldsymbol\pi\prime\}) =
\sum_{i=1}^{N}w_i\Delta M_i^2 p_i\prime = \sum_{i=1}^{N} w_i
(\mathfrak{f}(X_i,\{\boldsymbol\pi\})- M^{PDB})^2 p_i\prime
\end {equation}
with respect to $\{\boldsymbol\pi\}$. $p_i\prime$ is defined by
equation \ref{eq:pi}. Again, we have added the relative weights $w_i$ to
adjust for the effect of $\Lambda$,$\Sigma$ and $^{12}_\Lambda$B GS
events in the calibration. The new values of the parameter is used to
update the probabilities $p_i\prime$, and the $\mathfrak{g}$ function is again
minimized. Comparing with eq.\ref{eq:chis}, we can see that the EM algorithm in
this case is nothing but an iteratively reweighted least-square
procedure. The weights or probabilies are not constants now, but 
functions of $\Delta M_i$ or $\{\boldsymbol\pi\}$. Example probability functions calculated
for the HKS spectrometer calibration is shown in fig.\ref{fig:rw}.

\begin{figure}
\begin{center}
\includegraphics [width=12cm]{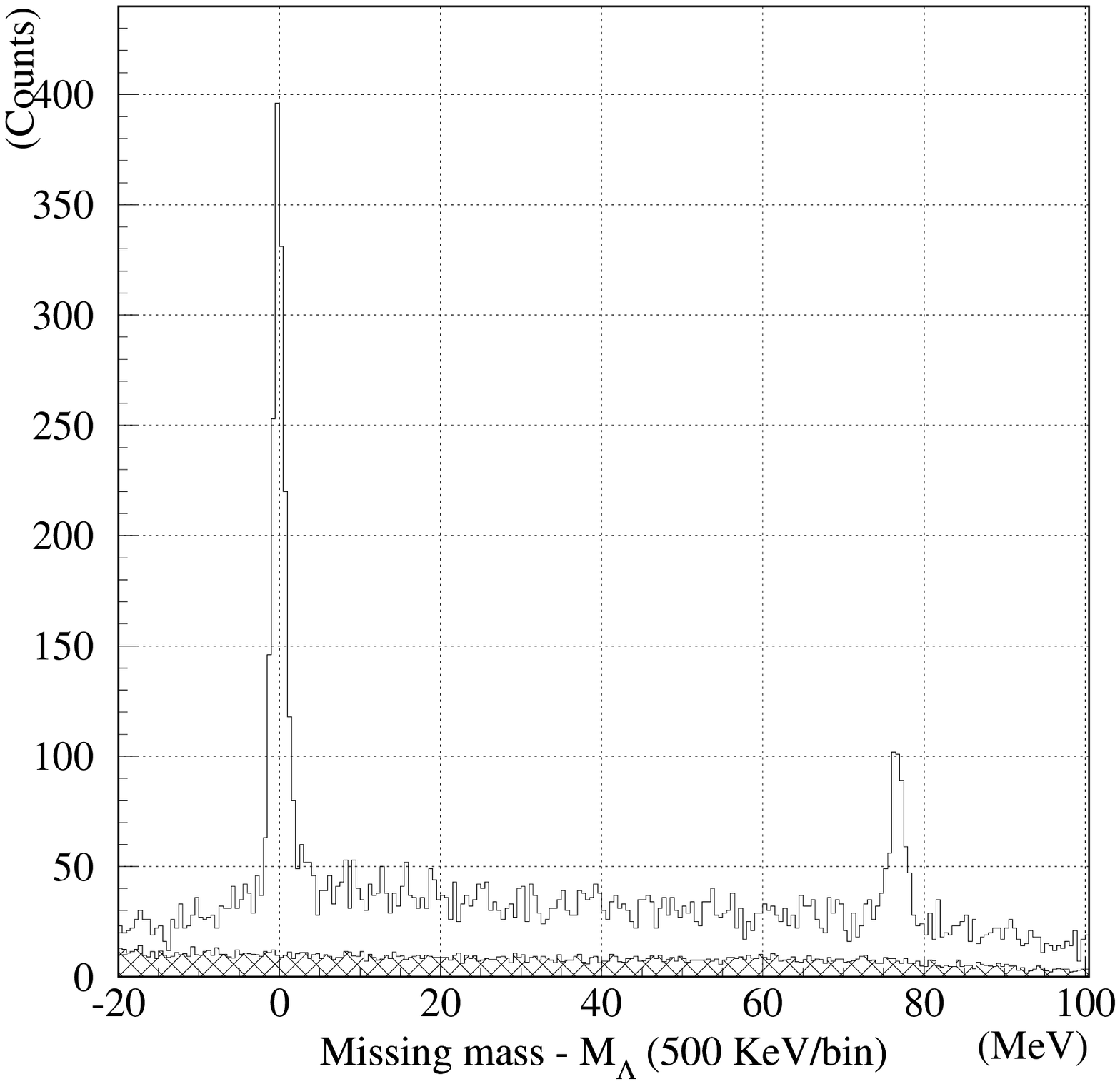}
\end{center}
\caption{$\Lambda$ and $\Sigma$ missing mass distribution produced by
  $p$(e,e'K) reaction from CH$_2$ target. The shaded region is
  accidental backgroud.}
\label{fig:ch2}
\end{figure}

The minimization of function $\mathfrak{g}$ is also carried out by CERNLIB
Fortran program package LEAMAX. We have obtained preliminary
reconstruction functions by the EM method described above. The
preliminary missing mass spectra of
$\Lambda$,$\Sigma^0$ from CH$_2$ target and hypernucleus
$^{12}_\Lambda$B from  C$^{12}$ are shown in fig. \ref{fig:c12} and
fig. \ref{fig:ch2} overlayed
with background.

 \end{document}